\documentclass[aps,pra,showpacs,superscriptaddress,10pt]{revtex4-2}

\usepackage[utf8]{inputenc}
\usepackage[english]{babel}

\usepackage{amssymb}
\usepackage{color}
\usepackage{amsmath}
\usepackage{graphicx}
\usepackage{bm}
\usepackage{braket}
\usepackage{booktabs}
\usepackage{multirow}
\usepackage{url}

\begin{document}

\title{Variational Quantum Eigensolver for the Analysis of High-Resolution NMR Spectra: Applications to AB and AB$_2$ Spin Systems}

\author{Murat Kurt}
\email{kmuratphysics@gmail.com}
\affiliation{Department of Software Engineering, Samsun University, 55420 Samsun, T\"{u}rkiye}

\author{Ayda Kaltehei}
\email{ayda.kaltehei@gmail.com}
\affiliation{Department of Physics, Ondokuz May{\i}s University, 55139 Samsun, T\"{u}rkiye}

\author{Sel\c{c}uk \c{C}akmak}
\email{selcuk.cakmak@samsun.edu.tr}
\affiliation{Department of Software Engineering, Samsun University, 55420 Samsun, T\"{u}rkiye}

\author{Azmi Gen\c{c}ten}
\email{gencten@omu.edu.tr}
\affiliation{Department of Physics, Ondokuz May{\i}s University, 55139 Samsun, T\"{u}rkiye}


\begin{abstract}
The Variational Quantum Algorithms (VQAs) are hybrid quantum-classical algorithms and
they can be used in the Nosiy Intermadiate Scale Quantum (NISQ) devises. The Variational
Quantum Eigensolver (VQE) was suggested as a first VQA. VQE is based on the variational
method of quantum mechanics and it is used to find the ground state energy of a quantum
system. In this study, VQE is used for the analysis of NMR spectra for the AB and AB$_2$ spin
systems. The frequencies and the spin coupling values are obtained from the sample spectra
for these spin systems. Then the Hamiltonians are written in terms of pauli spin operators and
transformed into a suitable forms for quantum computer. By employing VQE the ground state
energies are obtained for the related spin systems. They are found to be in good agreement with
the results obtained from the known variation method.

\vspace{10pt}
\begin{description}
\item[Keywords]NISQ, VQA, NMR, VQE

\end{description}
\end{abstract}

\maketitle

\section{\label{sec:intro} Introduction}
Quantum computers use qubits, which can be both 0 and 1 at the same time. This is the superposition property of quantum computers. Another significant property of quantum computers is entanglement, which plays a crucial role in enabling quantum correlations and information processing. Due to these features, quantum computers have more advantages than classical ones in terms of storage capacity and processing speed \cite{Benioff1980,Feynman1981,Steane1997}.Quantum information processing represents an advanced computational paradigm that leverages the fundamental principles of quantum mechanics to solve complex problems that are intractable for classical systems. In quantum computing, specific algorithms are employed based on the requirements of the task two well-known examples are Shor's algorithm for breaking down large numbers into prime factors and Grover's algorithm for finding items in a large, unsorted list \cite{Shor1996FTQC,Grover1996}. In recent years, Noisy Intermediate-Scala Quantum (NISQ) computers have started to be used \cite{Preskill2018,Bharti2022}. NISQ devices contain limited number of qubits and are not error corrected. These constraints hinder the implementation of large-scale quantum algorithms and elevate the importance of designing resource-efficient approaches. In response, hybrid quantum-classical algorithms, such as the Variational Quantum Algorithms (VQAs), have emerged as promising candidates \cite{Cerezo2020}. By offloading parts of the computation to classical processors and reducing quantum circuit depth, VQAs help alleviate the impact of noise and improve the feasibility of practical quantum applications on near-term hardware.\\

VQAs having hybrid quantum-classical properties can be used in NISQ devices \cite{Liu2021,Grange2023,farhi2014,Qi2024VQA,Huang2023,Chen2021HybridQC,McClean2016,Moll2018,Jones2019}. They are developed as an alternative to fully quantum algorithm such as quantum phase estimation which require foult-tolerant quantum hardware. The Variational Quantum Eigensolver (VQE) is one of the VQAs and it is used to find the ground state energy values for the selected problems \cite{Fedorov2022VQE,Tilly2022VQEReview}. VQE was first proposed by Peruzzo et al \cite{Peruzzo2014}.Then theoretical extension of VQE is presented by McClean et al \cite{McClean2016}. VQE has subsequently been applied to various problems and they are mentioned in recents review studies \cite{Fedorov2022VQE,Tilly2022VQEReview}. VQE is designed to solve Hamiltonian eigenvalue problems commonly encountered in fields such as quantum chemistry and materials science. Determining the ground state energies and eigenstates of quantum systems is a key step in addressing many problems within these domains. Some examples will be given as follows. VQE was applied for closed shell molecules in order to assess the ground state energies of H$_2$O,N$_2$ and Li$_2$O \cite{Kim2024NonBosonicVQE}. By using a four qubit photonic processor the ground state energy of He atom was obtained with VQE \cite{ghavami2025he}. Using VQE, the ground state energy of benzene was determined as a function of spatial deformation \cite{Sennane2023BenzeneVQE}. In order to evaluate the ground state energy of the antiferromagnetic Heisenberg model,  Large-scale simulations are performed by using a hybrid quantum–classical variational method \cite{Jattana2022HeisenbergVQE}. Simulating NMR spectra with a quantum computer is presented \cite{OssorioCastillo2024Simulating}. In this study bu using VQE the lowest eigenvalue for Sulfanol molecule is calculated.\\

The variational method is a widely used and effective computational technique for the approximate determination of ground and excited state energies of quantum systems. VQE approach is based on representing the system wave function with a set of adjustable parameters and obtaining the optimal approximation by minimizing the energy expectation value. In the analysis of high resolution NMR spectra the Hamiltonian of the spin systems typically contain resonance frequencies, the spin–spin coupling constants.  The VQE can be implemented on these Hamiltonians through parameterized quantum circuits. In this way, the ground state energy levels can be calculated with high accuracy. This approach plays a significant role both in the analysis of fundamental quantum mechanical problems and in the development of quantum information processing–based simulation techniques. The fundamental concept involves preparing a quantum state on a quantum computer via a parameterized quantum circuit, known as an ansatz, followed by measurement of this state which is fed into a classical optimization loop. This classical loop updates the circuit parameters to minimize a given objective function, thereby guiding the algorithm. This approach is well suited to current quantum NISQ due to its ability to operate with limited qubit depth~\cite{PerezRamirez2024}.\\

In this study, VQE is used for the Analysis of High Resolution NMR Spectra for the AB and A$B_2$ spin systems. For both spin systems resonance frequencies and spin coupling constants are determined by the spectra obtained from the literature. These data are served as the basis for constructing Hamiltonian operators describing the respective spin systems. Subsequently, these Hamiltonians were expressed in terms of Pauli matrices and transformed into a form suitable for execution on quantum computers. In the following step, quantum circuits compatible with the VQE algorithm were designed for both AB and AB$_2$ spin systems, and the ground state energies of each spin systems were computed. The calculated ground state energies were validated by comparison with the theoretical results obtained via variational methods. It was observed that the obtained energy values exhibited a high degree of agreement with theoretical expectations.

\section{\label{sec:theory} Theory}

\subsection{Quantum Information Processing}

Quantum information processing (QIP) is a computational model in which information is processed within the framework of the principles of quantum mechanics. In this model, the information carriers, called qubits, are represented by quantum states defined in a n-dimensional Hilbert space. The state of a single qubit can be expressed as

\begin{equation}
|\psi\rangle = \alpha |0\rangle + \beta |1\rangle
\end{equation}
\\
Here is $\quad |\alpha|^2+|\beta|^2=1$. Multi qubit systems are defined in the $\mathcal{H}^{\otimes n}$ Hilbert space, which is formed by the tensor product of single qubit states and the general state of a n-qubit system can be expressed as

\begin{equation}
|\psi\rangle = \sum_{j=0}^{2^n - 1} \alpha_j \, |j\rangle,
\quad
\sum_{j=0}^{2^n - 1} |\alpha_j|^2 = 1.
\end{equation}\\
Unlike classical probability distributions, this allows for the emergence of features such as quantum superposition and entanglement. In particular, entangled states enable the processing of correlations that are impossible in classical systems, forming the foundation of the superiority of quantum algorithms. The process of QIP is carried out through unitary transformations and measurement operators. Unitary operators $U$ determine the time evolution of quantum states and represent norm-preserving transformations in the Hilbert space.\\

In this study, instead of general unitary operations and measurement operators as described in the standard quantum algorithm framework, specific quantum gates will be employed to construct the quantum circuits. These gates include the NOT gate, the controlled-NOT (CNOT) gate, the rotation gate \( R(\theta) \), and the controlled rotation gate \( \text{C-R}(\theta) \). Each of these gates has distinct functionality and plays a key role in encoding, manipulating, and entangling qubits within the quantum circuit. NOT gate flips the state of a single qubit from \(|0\rangle\) to \(|1\rangle\) and vice versa. It is mathematically represented by the Pauli-X matrix

\begin{equation}
X = \begin{bmatrix}
0 & 1 \\
1 & 0
\end{bmatrix}
\end{equation}\\
The controlled-NOT gate operates on two qubits, applying the \(X\) gate to the target qubit only when the control qubit is in the state \(|1\rangle\). Its matrix form is

\begin{equation}
\mathrm{CNOT} =
\begin{bmatrix}
1 & 0 & 0 & 0 \\
0 & 1 & 0 & 0 \\
0 & 0 & 0 & 1 \\
0 & 0 & 1 & 0
\end{bmatrix}
\end{equation}\\
Single qubit rotation gate performs a rotation around a chosen axis (commonly Y or Z) on the Bloch sphere. A common Y-axis rotation form is

\begin{equation}
R_y(\theta) =
\begin{bmatrix}
\cos(\theta/2) & -\sin(\theta/2) \\
\sin(\theta/2) & \cos(\theta/2)
\end{bmatrix}
\end{equation}\\
Two qubits controlled rotation gate applies the rotation \( R(\theta) \) to the target qubit only if the control qubit is in state \(|1\rangle\). This gate is particularly useful in parametrized quantum circuits for encoding continuous variables or entangling operations.

\subsection{VQE}

The variational method in quantum mechanics provides an approximate way 
to determine the ground state energy of a system. For any trial wavefunction 
\(|\psi_{\text{trial}}\rangle\), the variational principle states is

\begin{equation}
E_0 \leq 
\frac{\langle \psi_{\text{trial}} | H | \psi_{\text{trial}} \rangle}
     {\langle \psi_{\text{trial}} | \psi_{\text{trial}} \rangle} .
\end{equation}
The variation method is used in VQE. In VQE a parametrized state \(|\psi(\boldsymbol{\theta})\rangle\) is introduced,
where \(\boldsymbol{\theta}\) denotes a set of variational parameters, 
one minimizes this expectation value with respect to \(\boldsymbol{\theta}\) 
to approximate the ground state energy. VQE aims to minimize a given cost function by combining the quantum computer's ability to efficiently explore high-dimensional Hilbert spaces with classical optimization techniques. This approach offers significant advantages, particularly in fields such as molecular energy calculations, optimization problems, and quantum machine learning. The core component of VQE is the parametrized quantum circuit. Starting from an initial state $|\psi_0\rangle$, a quantum state is prepared by applying a parametrized unitary transformation $U(\boldsymbol{\theta})$ called ansatz.\\
\begin{equation}
|\psi(\boldsymbol{\theta})\rangle = U(\boldsymbol{\theta})|\psi_0\rangle,
\end{equation}\\
where $\boldsymbol{\theta} = \{\theta_1, \theta_2, \dots, \theta_p\}$ represents the tunable parameters of the quantum gates in the circuit. Typically, $U(\boldsymbol{\theta})$ consists of a combination of single- and two-qubit gates, designed to explore a specific subspace of the Hilbert space.\\

In VQE, the objective is to approximate the ground-state energy of a target Hamiltonian $H$. The energy expectation value of the parametrized state is defined as

\begin{equation}
E(\boldsymbol{\theta}) = \langle \psi(\boldsymbol{\theta}) | H | \psi(\boldsymbol{\theta}) \rangle.
\end{equation}\\
During this measurement process, parameter updates are performed by a classical optimization algorithm. The operation of a VQE is a cyclic procedure that combines quantum and classical subroutines. These steps are shown in Figure 1.

\begin{enumerate}
    \item A parameter vector $\boldsymbol{\theta}$ is chosen, and the quantum circuit $U(\boldsymbol{\theta})$ is applied to prepare the state $|\psi(\boldsymbol{\theta})\rangle$.
    \item Expectation values for each Pauli term of the Hamiltonian are measured on the quantum computer.
    \item The total energy $E(\boldsymbol{\theta})$ is calculated, and the parameter vector $\boldsymbol{\theta}$ is updated using a classical optimization algorithm (e.g., Nelder-Mead, COBYLA, or gradient-based methods).
    \item Steps 1–3 are repeated until a predefined convergence criterion is satisfied.
\end{enumerate}

\begin{figure}[htbp]
  \centering
  \includegraphics[width=0.7\linewidth]{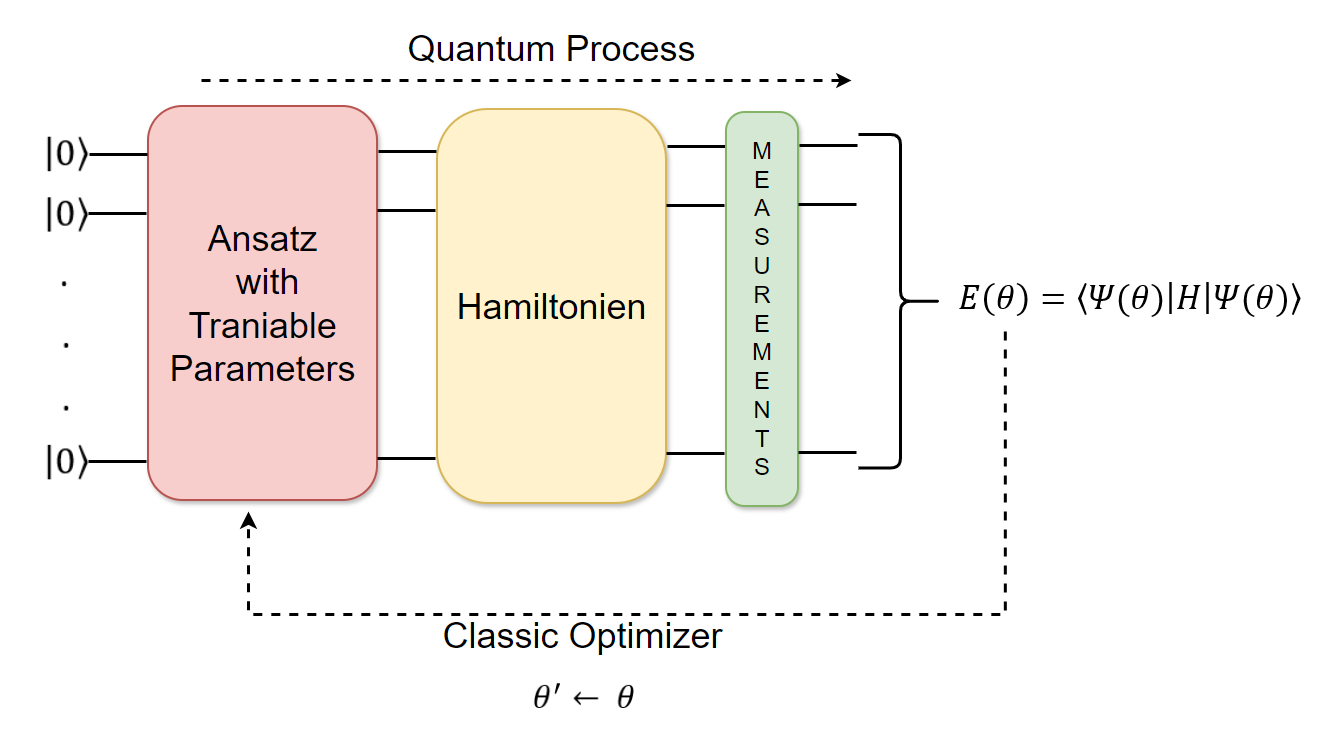}
  \caption{The VQE workflow.}
  \label{fig:vqa_workflow}
\end{figure}

\subsection{The Variational Method in NMR}

In NMR, the general Hamiltonian is given by
\begin{equation}
H = -\sum_{i} \nu_i I_z(i) + \sum_{i<j} J_{ij} \, \vec{I}(i) \cdot \vec{I}(j),
\end{equation}
where $\nu_i$ denotes the Larmor frequency of the $i$-th nucleus in the magnetic field, and $J_{ij}$ represents the spin–spin coupling constant (J-coupling) between nucleies $i$ and $j$. Using quantities obtained from NMR spectra, the spin Hamiltonians can be constructed for the releated spin systems.  
In this study, AB and A$B_2$ spin systems were analyzed, and the computational relations for these two systems are provided below.\\

The hamiltonien for the AB spin system can be expressed as

\begin{equation}
H = -\nu_A I_{zA} 
    -\nu_B I_{zB} 
    + J_{AB} \, \vec{I}_A \cdot \vec{I}_B.
\end{equation}
For AB spin system, we have four states: $\displaystyle |\alpha\alpha\rangle$,$\displaystyle |\alpha\beta\rangle$, $\displaystyle |\beta\alpha\rangle$ and $\displaystyle |\beta\beta\rangle$. Then the Hamiltonien matrix for AB spin system can be obtained as (J = J$_{AB}$) \\ 

\begin{equation}
\hat{H}_{AB} =
\begin{bmatrix}
 = -\frac{\nu}{2}-\frac{\nu}{2}+\frac{J}{4} & 0 & 0 & 0 \\
0 & -\frac{\nu}{2}+\frac{\nu}{2}-\frac{J}{4} & \frac{J}{2} & 0 \\
0 & \frac{J}{2} & \frac{\nu}{2}-\frac{\nu}{2}-\frac{J}{4} & 0 \\
0 & 0 & 0 & \frac{\nu}{2}+\frac{\nu}{2}+\frac{J}{4}
\end{bmatrix}
\end{equation}
By using the variational method, the eigenvalues and eigenstates can be found as given in Table 1. A detailed analysis can be found in the literature \cite{gunther2023nmr,abraham1999introduction}. 

\begin{table}[h]
\caption{ The eigenstates and eigenvalues for the AB spin system \cite{gunther2023nmr,abraham1999introduction}.}
\label{tab:AB_eigenstates}
\begin{tabular}{@{}lll@{}}
\toprule
\textbf{State} & \textbf{Dirac Representation} & \textbf{Eigenvalue $E_i$} \\
\midrule
$\displaystyle |\alpha\alpha\rangle$
& $\displaystyle |00\rangle$
& $\displaystyle -\frac{1}{2}(\nu_A+\nu_B)+\frac{J}{2}$ \\

$\displaystyle \cos\theta\,|\alpha\beta\rangle + \sin\theta\,|\beta\alpha\rangle$
& $\displaystyle \cos\theta\,|01\rangle + \sin\theta\,|10\rangle$
& $\displaystyle -\frac{J}{4}-C$ \\

$\displaystyle -\sin\theta\,|\alpha\beta\rangle + \cos\theta\,|\beta\alpha\rangle$
& $\displaystyle -\sin\theta\,|01\rangle + \cos\theta\,|10\rangle$
& $\displaystyle -\frac{J}{4}+C$ \\

$\displaystyle |\beta\beta\rangle$
& $\displaystyle |11\rangle$
& $\displaystyle \frac{1}{2}(\nu_A+\nu_B)+\frac{J}{2}$ \\
\botrule
\end{tabular}
\end{table}

For AB spin system, the transition frequencies $f_1$, $f_2$, $f_3$, and $f_4$ are obtained from the spectrum. These values are then used to solve the following set of equations to determine $\nu_A$, $\nu_B$, and $J$ values:

\begin{equation}
f_3 - f_4 = f_1 - f_2 = J,
\end{equation}
\begin{equation}
f_2 - f_4 = f_1 - f_3 = 2C,
\end{equation}
\begin{equation}
\frac{f_1 - f_3}{2} = \frac{f_2 - f_4}{2} = C,
\end{equation}
\begin{equation}
f_1 + f_4 = \nu_A + \nu_B,
\end{equation}
\begin{equation}
C = \frac{1}{2} \sqrt{J^2 + (\nu_A - \nu_B)^2},
\end{equation}
\begin{equation}
\nu_A - \nu_B = \sqrt{4C^2 - J^2}.
\end{equation}
The value of $\theta$ required to obtain the excited states in the AB spin system is calculated as:
\begin{equation}
\tan(2\theta) = \frac{J}{\nu_A - \nu_B}.
\end{equation}\\

The hamiltonien for the AB$_2$ spin system can be expressed as

\begin{equation}
H = -\nu_A I_z(A) - \nu_{B1} I_z(B1) - \nu_{B2} I_z(B2) 
    + J_{AB(1)} \, \vec{I}(A) \cdot \vec{I}(B1) 
    + J_{AB(2)} \, \vec{I}(A) \cdot \vec{I}(B2).
\end{equation}\\
For B$_2$ spin system by using the variation method $|\alpha\alpha\rangle$,$\; \frac{1}{\sqrt{2}}\left(|\alpha\beta\rangle + |\beta\alpha\rangle\right)$, $\frac{1}{\sqrt{2}}\left(|\alpha\beta\rangle - |\beta\alpha\rangle\right)$, $|\beta\beta\rangle$ states can be found \cite{gunther2023nmr}. As A nucleus have $|\alpha\rangle$ and $|\beta\rangle$ states, eight basis functions for AB$_2$ spin system can be written as given in Table 2. Then, the Hamiltonien matrix for AB$_2$ spin system is obtained as ($J=J_{AB}$) \\

\begin{equation}
\hat{H}_{AB_2}=
\begin{bmatrix}
-\frac{\nu_A}{2}-\nu_B+\frac{J}{2} & 0 & 0 & 0 & 0 & 0 & 0 & 0\\
0 & -\frac{\nu_A}{2} & 0 & 0 & 0 & 0 & 0 & 0\\
0 & 0 & -\frac{\nu_A}{2} & \frac{J}{\sqrt{2}} & 0 & 0 & 0 & 0\\
0 & 0 & \frac{J}{\sqrt{2}} & \frac{\nu_A}{2}-\nu_B-\frac{J}{2} & 0 & 0 & 0 & 0\\
0 & 0 & 0 & 0 & -\frac{\nu_A}{2}+\nu_B-\frac{J}{2} & \frac{J}{\sqrt{2}} & 0 & 0\\
0 & 0 & 0 & 0 & \frac{J}{\sqrt{2}} & \frac{\nu_A}{2} & 0 & 0\\
0 & 0 & 0 & 0 & 0 & 0 & \frac{\nu_A}{2} & 0\\
0 & 0 & 0 & 0 & 0 & 0 & 0 & \frac{\nu_A}{2}+\nu_B+\frac{J}{2}
\end{bmatrix}
\end{equation}\\

\begin{table}[h]
\caption{The basis functions for AB$_2$ spin system.}
\label{tab:AB_eigenstates_mT}
\begin{tabular}{@{}c l l@{}}
\toprule
\textbf{$m_T$} & \textbf{State} & \textbf{Dirac Representation} \\
\midrule
$\displaystyle \frac{3}{2}$
& $\displaystyle |\alpha\alpha\alpha\rangle$
& $\displaystyle |000\rangle$ \\
\hline
\multirow{3}{*}{$\displaystyle \frac{1}{2}$}
& $\displaystyle \frac{1}{\sqrt{2}}\,(|\alpha\alpha\beta\rangle -|\alpha\beta\alpha\rangle)$
& $\displaystyle \frac{1}{\sqrt{2}}\,(|001\rangle -|010\rangle)$\\

& $\displaystyle \frac{1}{\sqrt{2}}\,(|\alpha\alpha\beta\rangle +|\alpha\beta\alpha\rangle)$
& $\displaystyle \frac{1}{\sqrt{2}}\,(|001\rangle +|010\rangle)$\\

& $\displaystyle |\beta\alpha\alpha\rangle $
& $\displaystyle |100\rangle$ \\
\hline
\multirow{3}{*}{$\displaystyle -\frac{1}{2}$}
& $\displaystyle |\alpha\beta\beta\rangle$
& $\displaystyle |011\rangle$ \\

& $\displaystyle \frac{1}{\sqrt{2}}\,(|\beta\alpha\beta\rangle +|\beta\beta\alpha\rangle)$
& $\displaystyle \frac{1}{\sqrt{2}}\,(|101\rangle +|110\rangle)$\\

& $\displaystyle \frac{1}{\sqrt{2}}\,(|\beta\alpha\beta\rangle -|\beta\beta\alpha\rangle)$
& $\displaystyle \frac{1}{\sqrt{2}}\,(|101\rangle -|110\rangle)$\\
\hline
$\displaystyle -\frac{3}{2}$
& $\displaystyle |\beta\beta\beta\rangle$
& $\displaystyle |111\rangle$ \\
\botrule
\end{tabular}
\end{table}

By using the variational method the eigenvalues and eigenstates can be found as given in Table 3. A more detailed analysis can be found in the literature \cite{gunther2023nmr,abraham1999introduction,OssorioCastillo2024Simulating}.

\begin{table}[h]
\caption{The eigenstates and corresponding the energy values for the AB$_2$ spin system.}
\label{tab:AB_eigenstates_energy}
\begin{tabular}{@{}l c@{}}
\toprule
\textbf{Eigen states} & \textbf{Energy values} \\
\midrule
$|\alpha\alpha\alpha\rangle$ 
& $-\frac{\nu_A}{2}-\nu_B+\frac{J}{2}$ \\

$\dfrac{cos\theta_+}{\sqrt{2}}\,(|\alpha\alpha\beta\rangle +|\alpha\beta\alpha\rangle)+sin\theta_+|\beta\alpha\alpha\rangle$
& $\frac{1}{2}(-\nu_B-\frac{J}{2})+C_+$ \\

$-\dfrac{sin\theta_+}{\sqrt{2}}\,(|\alpha\alpha\beta\rangle +|\alpha\beta\alpha\rangle)+cos\theta_+|\beta\alpha\alpha\rangle$
& $\frac{1}{2}(-\nu_B-\frac{J}{2})-C_+$ \\

$\dfrac{1}{\sqrt{2}}\,(|\alpha\alpha\beta\rangle +|\alpha\beta\alpha\rangle)$
& $-\frac{\nu_A}{2}$ \\

$cos\theta_-|\alpha\beta\beta\rangle+\dfrac{sin\theta_-}{\sqrt{2}}\,(|\beta\alpha\beta\rangle +|\beta\beta\alpha\rangle)$
& $\frac{1}{2}(\nu_B-\frac{J}{2})+C_-$ \\

$-sin\theta_-|\alpha\beta\beta\rangle+\dfrac{cos\theta_-}{\sqrt{2}}\,(|\beta\alpha\beta\rangle +|\beta\beta\alpha\rangle)$
& $\frac{1}{2}(\nu_B-\frac{J}{2})-C_-$ \\

$\dfrac{1}{\sqrt{2}}\,(|\beta\alpha\beta\rangle +|\beta\beta\alpha\rangle)$
& $\frac{\nu_A}{2}$ \\

$|\beta\beta\beta\rangle$
& $\frac{\nu_A}{2}+\nu_B+\frac{J}{2}$ \\
\botrule
\end{tabular}

\vspace{2mm}
\footnotesize{
Here, $ tan2\theta_+=\frac{\sqrt{2}J}{\nu_A-\nu_B-\frac{J}{2}}$,$ tan2\theta_-=\frac{\sqrt{2}J}{\nu_A-\nu_B+\frac{J}{2}}$ and $C_+=\frac{1}{2}\sqrt{(\nu_A-\nu_B-\frac{J}{2})^2+2J^2}$, $C_-=\frac{1}{2}\sqrt{(\nu_A-\nu_B+\frac{J}{2})^2+2J^2}$
}
\end{table}

For this spin system experimentaly there exist maximum nine transitions. But in a spectrum of an AB$_2$ spin system there are eight evident lines. The values $f_1, f_2, f_3, f_4, f_5, f_6, f_7$, and $f_8$ are used in the following equations to determine $\nu_A$, $\nu_B$, and $J_{AB}$ values

\begin{equation}
    \nu_A = f_3,
\end{equation}
\begin{equation}
    \nu_B = \frac{f_5 + f_7}{2},
\end{equation}
\begin{equation}
    J_{AB} = \frac{(f_1 - f_4) + (f_6 - f_8)}{3}.
\end{equation}

\section{\label{sec:results} Results and Discussion}

\subsection{VQE for AB Spin System}

For an AB spin system, the Hamiltonian can be expressed in terms of Pauli spin operators as follows,

\begin{equation}
H = 
-\frac{\nu_A}{2}\,(Z \otimes I)
-\frac{\nu_B}{2}\,(I \otimes Z)
+ \frac{J_{AB}}{4}\,\bigl(
X \otimes X \;+\; 
Y \otimes Y \;+\;
Z \otimes Z
\bigr)
\end{equation}
For this spin system, the ${^1H}$ NMR spectrum of 2,4-dibromothiophene recorded at 300 MHz was used~\cite{chemicalbook_24dibromo_nmr}. In this spectrum the measured frequencies were $f_1 = 2094.84$ Hz, $f_2 = 2093.20$ Hz, $f_3 = 2061.80$ Hz, and $f_4 = 2060.16$ Hz. Using these values and applying Equations (12), (13), (14), (15), (16) and (17), the resonance frequencies and the coupling constants were calculated as $\nu_A = 2094.007$ Hz and $\nu_B = 2060.99$ Hz and $J_{AB} = 1.64$ Hz. By using this Hamiltonian given by Equation 24, the ground-state energy of the AB spin system can be calculated via VQE.\\

\begin{figure}[htbp]
  \centering
  \includegraphics[width=0.7\linewidth]{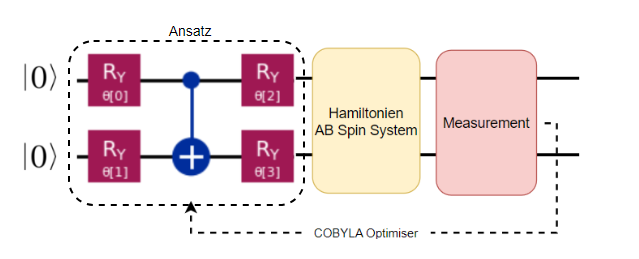}
  \caption{The full VQE process with ansatz part for the AB spin system.}
  \label{fig:ab_ansatz}
\end{figure}

For the determination of the ground state energy of the AB spin system, a two-qubit quantum circuit was designed. The first qubit represents the spin state of nucleus A, while the second qubit represents the spin state of nucleus B. Since the ground state is considered, both qubits are initialized to $\quad|00\rangle$. The ansatz circuit, which serves as the trial wavefunction, consists of four rotation gates and one CNOT gate. The initial values of the rotation angles were set to 1. The full VQE process with ansatz part is shown in Figure~\ref{fig:ab_ansatz}.

The measurement outcomes obtained at each step were processed using the COBYLA optimization method to obtain the ground state. The optimized parameters were found to be:

\[
\theta_{1} = 5.488 \times 10^{-5}, \quad
\theta_{2} = 1.032 \times 10^{-1}, \quad
\theta_{3} = -4.261 \times 10^{-5}, \quad
\theta_{4} = -1.031 \times 10^{-1}.
\]

\begin{figure}[htbp]
  \centering
  \includegraphics[width=0.7\linewidth]{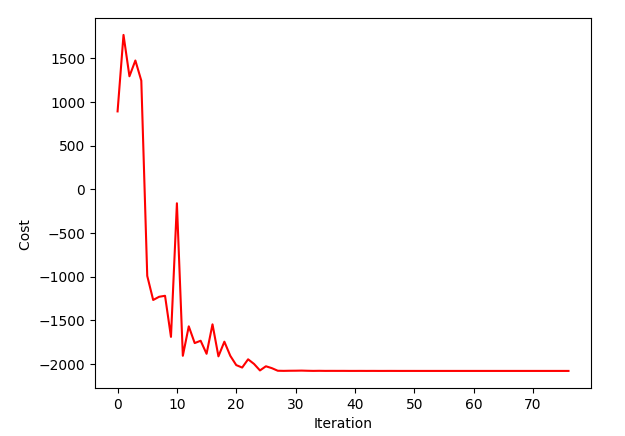}
  \caption{Cost function for the VQE circuit of the AB spin system.}
  \label{fig:ab_cost}
\end{figure}

The ground state energy was calculated as $-2077.907$ Hz. Figure~\ref{fig:ab_cost} shows the cost function versus iteration plot for the VQE circuit used in this calculation. The ground state energy from the known variational method is obtained as $\displaystyle -\frac{1}{2}(\nu_A+\nu_B)+\frac{J}{2} = -2081.178 Hz$. So it is found that the both results are a good aggrement.

\subsection{VQE for AB$_2$ Spin System}

The Hamiltonian operators for the AB$_2$ spin system, expressed in terms of Pauli spin operators is given by

\begin{equation}
\begin{aligned}
H &= -\frac{\nu_A}{2} \,(Z \otimes I \otimes I) 
     -\frac{\nu_B}{2} \,(I \otimes Z \otimes I) 
     -\frac{\nu_B}{2} \,(I \otimes I \otimes Z) \\
  &\quad + \frac{J_{AB}}{4} \, \bigl( X \otimes X \otimes I + Y \otimes Y \otimes I + Z \otimes Z \otimes I \bigr) \\
  &\quad + \frac{J_{AB}}{4} \, \bigl( X \otimes I \otimes X + Y \otimes I \otimes Y + Z \otimes I \otimes Z \bigr)
\end{aligned}
\end{equation}

\begin{figure}[htbp]
  \centering
  \includegraphics[width=0.8\linewidth]{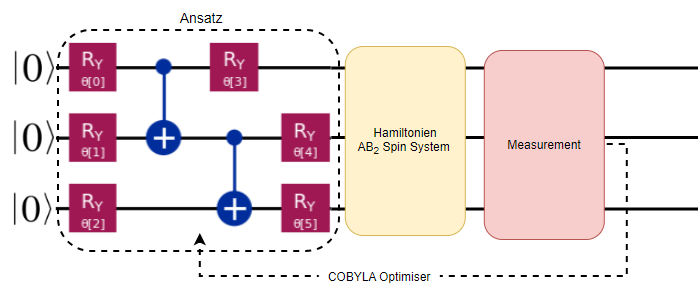}
  \caption{The full VQE process with ansatz part for the AB$_2$ spin system.}
  \label{fig:ansatz}
\end{figure}

For this spin system, the ${^1H}$ NMR spectrum of 2,6-dichlorobenzonitrile recorded at 200 MHz was used~\cite{reich_26dichloro_nmr}. In this spectrum the measured frequencies for eight lines are $f_1 = 1502.9$ Hz, $f_2 = 1498.0$ Hz, $f_3 = 1492.6$ Hz, $f_4 = 1487.8$ Hz, $f_5 = 1484.6$ Hz, $f_6 = 1484.2$ Hz, $f_7 = 1479.1$ Hz, and $f_8 = 1474.4$ Hz. Using these values and applying Equations (21), (22), and (23), the frequancies and the coupling constant were determined as $\nu_A = 1492.6$ Hz ,$\nu_{B1} = \nu_{B2} = 1481.84$ Hz and $J_{AB} = 8.2$ Hz. In order to determine the ground state energy of the AB$_2$ spin system, a three-qubit full VQE process with ansatz part was designed as given in Figure~\ref{fig:ansatz}.
The first qubit represents the spin state of nucleus A, the second qubit represents the spin state of nucleus B(1), 
and the third qubit represents the spin state of nucleus B(2). 
Since the ground state is considered, all three qubits are initialized to the $\quad|000\rangle$ state. This ansatz part of this circuit represents the trial wavefunction, consists of six rotation gates and two CNOT gates. 
The initial values of the rotation angles were set to 1.

The measurement outcomes obtained at each step were processed using the COBYLA optimization method to obtain the minimum energy, i.e., the ground state. 
The optimized parameters were found to be:

\[
\begin{aligned}
\theta_{1} &= -9.367 \times 10^{-5}, \quad
\theta_{2} = 1.009 \times 10^{-5}, \quad
\theta_{3} = 2.500 \times 10^{-1}, \\
\theta_{4} &= 3.410 \times 10^{-5}, \quad
\theta_{5} = -4.240 \times 10^{-5}, \quad
\theta_{6} = -3.500 \times 10^{-1}.
\end{aligned}
\]

\begin{figure}[htbp]
  \centering
  \includegraphics[width=0.7\linewidth]{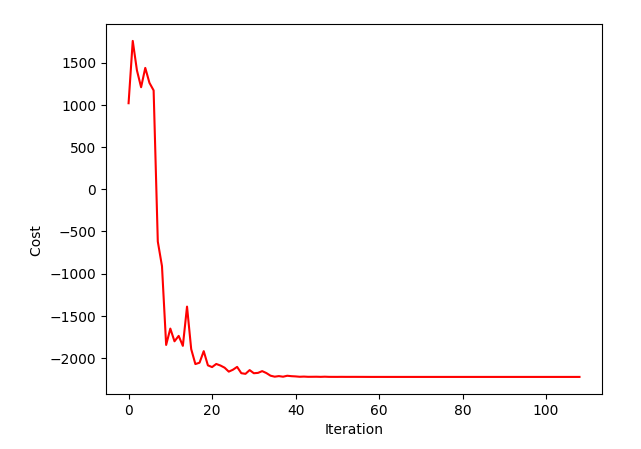}
  \caption{Cost function for the VQE circuit of the AB$_2$ spin system.}
  \label{fig:ab2_cost}
\end{figure}

So by using VQE the ground state energy was calculated as $-2224.03$ Hz. 
Figure~\ref{fig:ab2_cost} shows the cost as a function of iteration plot for the VQE circuit used in this calculation. By using the variational method the ground state energy from $-\frac{\nu_A}{2}-\nu_B+\frac{J}{2}$ is obtained as $-2224.04$ Hz showing a good agreement with the VQE value.

\section{\label{sec:conclusion} Conclusion}

VQAs are hybrid approaches that combine quantum and classical computation, making them suitable for implementation on NISQ devices. Among these methods, the VQE was proposed as one of the earliest VQAs. It relies on the variational principle of quantum mechanics and is commonly used to determine the ground-state energy of quantum systems. In this work, VQE is applied to the analysis of NMR spectra of AB and AB$_2$ spin systems. The resonance frequencies and spin–spin coupling parameters are extracted from sample spectra, after which the corresponding Hamiltonians are expressed using Pauli spin operators and reformulated in a quantum-computer–compatible representation. Using VQE, the ground-state energies of the considered spin systems are calculated and shown to be in strong agreement with results obtained via the conventional variational method. The results demonstrate that the VQE algorithm is an effective approach for accurately predicting ground state energies in NMR spin systems. This work reinforces the potential of quantum algorithms for simulating molecular spin systems and provides a foundation for exploring more complex multi-spin networks.

\bibliography{references} 
\bibliographystyle{apsrev4-2} 

\end{document}